\documentclass{jfm}

\usepackage{graphicx}
\usepackage{newtxtext}
\usepackage[stix2]{newtxmath}
\usepackage{natbib}
\usepackage[hidelinks]{hyperref}
\newcommand{\RomanNumeralCaps}[1]
\linenumbers
\usepackage{siunitx}

\def\bu{\boldsymbol{u}}
\def\del{\mathbf{\nabla}}

\def\pd{\partial}
\def\eps{\varepsilon}
\def\d{\mathrm{d}}

\newcommand*{\GL}{\hyperlink{cite.grossmann_scaling_2000}{{GL}}}

\newcommand\Nuss{\mbox{\textit{Nu}}}
\newcommand\Ray{\mbox{\textit{Ra}}}
\newcommand\Gr{\mbox{\textit{Gr}}}

\newcommand{\edit}{\textcolor{black}}

\graphicspath{{Figs/}}

\title{Boundary layers in turbulent vertical convection at high Prandtl number}

\author{Christopher J. Howland\aff{1}
  \corresp{\email{\href{mailto:c.j.howland@outlook.com}{c.j.howland@outlook.com}}},
  Chong Shen Ng\aff{1},
  Roberto Verzicco\aff{1,2,3}
 \and Detlef Lohse\aff{1,4}}

\affiliation{\aff{1}Physics of Fluids Group, Max Planck Center for Complex Fluid Dynamics, MESA+ Institute and J.\ M.\ Burgers Centre for Fluid Dynamics, University of Twente, P.O.\ Box 217, 7500AE Enschede, Netherlands
\aff{2}Dipartimento di Ingegneria Industriale, University of Rome ``Tor Vergata", Via del Politecnico 1, Roma 00133, Italy
\aff{3}Gran Sasso Science Institute - Viale F. Crispi, 7, 67100 L'Aquila, Italy
\aff{4}Max Planck Institute for Dynamics and Self-Organization, Am Fassberg 17, 37077 Göttingen, Germany
}

\begin{document}
\maketitle

\begin{abstract}
  Many environmental flows arise due to natural convection at a vertical surface, from flows in buildings to dissolving ice faces at marine-terminating glaciers.
  We use three-dimensional direct numerical simulations of a vertical channel with differentially heated walls to investigate such convective, turbulent boundary layers.
  Through the implementation of a multiple-resolution technique, we are able to perform simulations at a wide range of Prandtl numbers $\Pran$.
  This allows us to distinguish the parameter dependences of the horizontal heat flux and the boundary layer widths in terms of the Rayleigh number $\Ray$ and Prandtl number $\Pran$.
  For the considered parameter range $1\leq \Pran \leq 100$, $10^6 \leq \Ray \leq 10^9$, we find the flow to be consistent with a `buoyancy-controlled' regime where the heat flux is independent of the wall separation.
  For given $\Pran$, the heat flux is found to scale linearly with the friction velocity $V_\ast$.
  Finally, we discuss the implications of our results for the parameterisation of heat and salt fluxes at \edit{vertical} ice-ocean interfaces.
\end{abstract}


\vspace{-6ex}

\section{Introduction}
\label{sec:intro}

When a fluid is heated from a side boundary, buoyancy drives a flow up the boundary via convection.
The laminar flow along a heated surface has long been understood \citep{batchelor_heat_1954,kuiken_asymptotic_1968,shishkina_momentum_2016} but there is no formal solution for the case where the flow becomes turbulent.
This occurs when the Rayleigh number of the flow is sufficiently high.
In many environmental applications of this so-called \emph{vertical convection} (VC), such as the flow in a cavity wall, high Rayleigh numbers imply that an accurate understanding of the turbulent flow is needed to describe the heat transfer to the environment.

Such convective boundary layers are not only generated by surface heating.
For example, a vertical ice face submerged in salty water will drive convection due to the generation of fresh meltwater at the ice-water interface as it melts or dissolves \citep{mcconnochie_turbulent_2015,malyarenko_synthesis_2020}.
In this case, the buoyancy driving the flow is primarily due to the salinity difference between the meltwater and the ambient water.
One key difference between the two applications mentioned so far is the ratio of the diffusivities of momentum and heat (or salt), known as the Prandtl (or Schmidt) number $\Pran$.
In air the Prandtl number is $\Pran\approx 0.7$, whereas for salt diffusion in cold water the relevant parameter is $\Pran\approx 2000$.

Numerical simulations are often restricted to $\Pran=O(1)$ because high spatial resolution is needed at high $\Pran$ to resolve sharp scalar gradients that diffuse more slowly than the velocity gradients.
However, understanding the role of the Prandtl number is vital for interpreting the results of such research for environmental or geophysical applications.
We shall therefore investigate boundary layers in turbulent vertical convection at $\Pran\gg 1$ with the aim of bridging the gap from classical studies of convection to geophysical applications.

In this study, we use direct numerical simulations to investigate turbulent convective boundary layers for a range of Rayleigh and Prandtl numbers.
By using the multiple-resolution technique of \citet{ostilla-monico_multiple-resolution_2015}, we can efficiently simulate flows at high $\Pran$, and we vary $\Pran$ from 1 to 100.
Although this is still considerably lower than the $\Pran\approx 2000$ applicable to salt diffusion in the ocean, it is large enough to extract scaling laws in the large $\Pran$ regime, which we expect to also hold in oceanographic flows.

Many different setups have been used to investigate vertical convection boundary layers in numerical studies.
\citet{wang_regime_2021} recently simulated vertical convection in a closed box, but the presence of walls in that domain means that turbulent boundary layers are only observed at very high $\Ray$, at which only 2-D simulations are computationally feasible.
We instead simulate the flow in a vertical channel with periodic boundary conditions in the wall-parallel directions.
As originally described by \citet{batchelor_heat_1954}, this domain approximates the flow at mid-heights in a tall vertical cell.
A recent study by \citet{ke_law_2020} used this domain to simulate the temporally evolving boundary layer at a single heated wall, but in order to obtain converged statistics for a wide range of parameters, we instead consider the vertical channel setup where one wall is heated and the other is cooled.
This flow configuration achieves a statistically steady state with an anti-symmetric velocity profile and has been the subject of numerous numerical studies at $\Pran=O(1)$ \citep[e.g.][]{versteegh_direct_1999,pallares_turbulent_2010,ng_vertical_2015}.

The remainder of this paper is organised as follows.
In \S\ref{sec:sims} we outline the numerical model and the setup of the simulations.
This is followed by flow visualisations in \S\ref{sec:viz} and a qualitative discussion of $\Pran$-dependence of this flow.
In \S\ref{sec:heat_flux} we describe how various parameterisations for turbulent heat flux perform when applied to our simulations, and in \S\ref{sec:BLs} we identify appropriate scaling laws for the boundary layer thicknesses.
Finally, we conclude and discuss important remaining open questions for convective boundary layers in VC in \S\ref{sec:discussion}.
\edit{
  The paper is supplemented by a concrete translation of our results into the geophysical context, focusing on the transition from laminar-type to turbulent-type boundary layers (appendix \ref{app:BL_trans}) and a detailed analysis of the energy dissipation and thermal dissipation budgets.
}

\section{Numerical setup, simulations, and control and response parameters} \label{sec:sims}

\begin{figure}
  \centerline{\includegraphics[width=\linewidth]{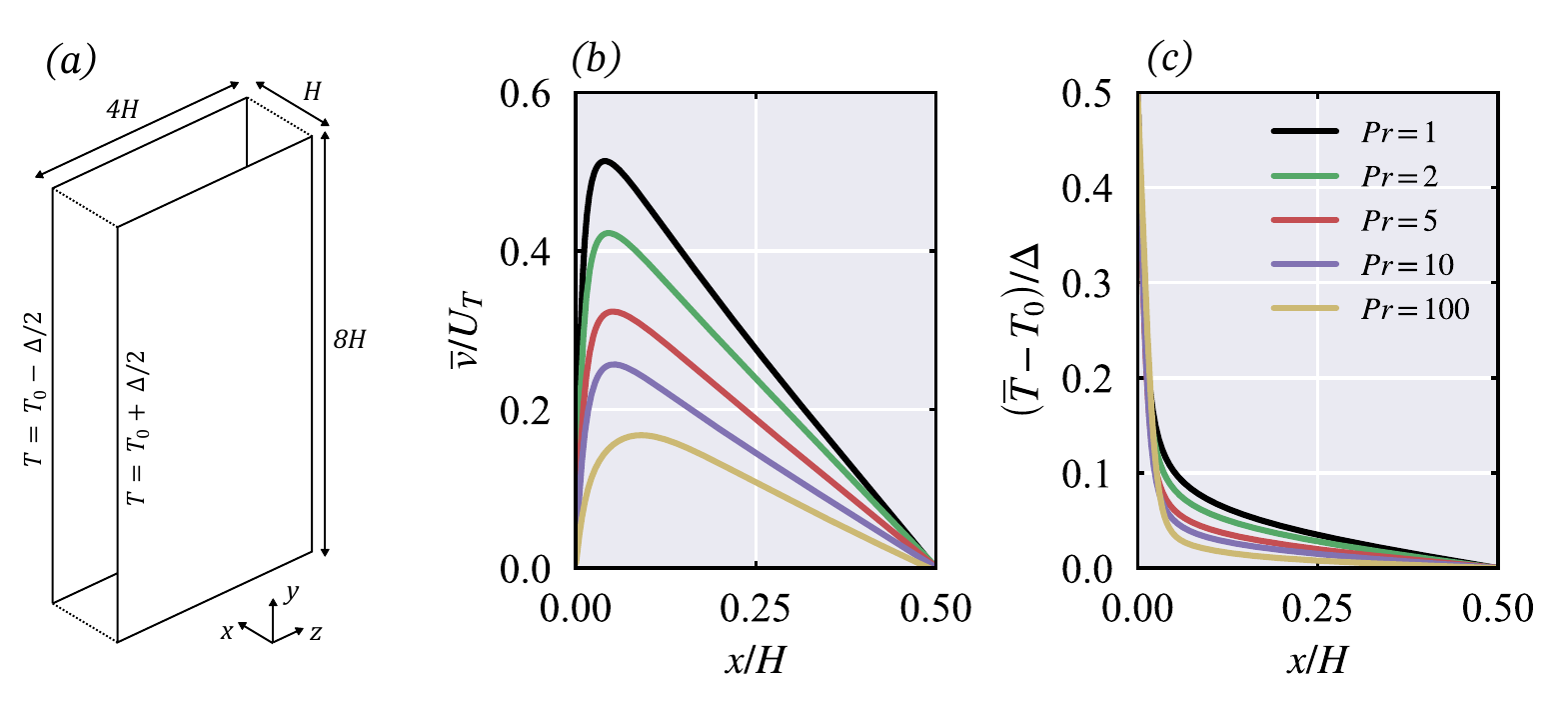}}
  \caption{
    $(a)$ A schematic of the simulation domain.
    $(b)$-$(c)$ Mean profiles of the vertical velocity and the temperature for $\Ray=10^8$ and a range of $\Pran$.
    Recall that the mean profiles are anti-symmetric such that $\overline{v}(x) = -\overline{v}(H-x)$.
  }
\label{fig:mean_profs}
\end{figure}

\edit{
  \subsection{Dynamical equations and control parameters}
}
We consider the Navier--Stokes equations subject to the Oberbeck--Boussinesq approximation, where changes in density $\rho$ are only relevant in the buoyancy and a linear equation of state relates the density changes to temperature $T$.
These equations read $\del \cdot \bu = 0$ and
\begin{align}
    \pd_t \bu + (\bu \cdot \del) \bu &= -\frac{1}{\rho_0}\del p + \nu \nabla^2 \bu + g \alpha T\hat{\mathbf{y}}, \label{eq:NS1}\\
    \pd_t T + \bu \cdot \del T &= \kappa \nabla^2 T,\label{eq:NS2}
\end{align}
where $\bu=(u,v,w)$ is the velocity field, $p$ the kinematic pressure, $\nu$ kinematic viscosity, $\kappa$ the molecular diffusivity of heat, $g$ gravitational acceleration, $\alpha$ the thermal expansion coefficient, and $\rho_0$ a reference density.
We solve these equations in a vertical channel domain between two no-slip, impermeable, isothermal walls.
These walls are separated by a distance $H$ and the temperature difference between them is $\Delta$.
As in \citet{ng_vertical_2015} and shown in figure \ref{fig:mean_profs}, we consider a domain of length $8H$ in the vertical ($y$) and length $4H$ in the spanwise ($z$) direction, and impose periodic boundary conditions on $\bu$, $p$, and $T$ in these directions, $y$ and $z$.
In a convective system, we can scale the velocity by the free-fall velocity $U_T=\sqrt{g\alpha \Delta H}$ so that the dynamics of the system are solely determined by the Rayleigh and Prandtl numbers
\begin{align}
    \Ray &= \frac{g\alpha H^3 \Delta}{\nu\kappa} , &
    \Pran &= \frac{\nu}{\kappa} . \label{eq:control_def}
\end{align}
These are the only control parameters of the system, aside from parameters characterising the geometry of the flow domain.
\edit{
  Their ratio $\Gr=\Ray/\Pran=g\alpha H^3 \Delta/\nu^2$ is also called the Grashof number.
}

\begin{table}
  \begin{center}
\def~{\hphantom{0}}
  \begin{tabular}{cccc}
      Prandtl number $\Pran$ & Rayleigh numbers $\Ray$ &  Max.\ base resolution & Max.\ scalar resolution  \\[3pt]
       1   & $10^6$ - $10^8$ & $384 \times 1536 \times 768$ & $768 \times 3072 \times 1536$\\
       2   & $10^6$ - $10^8$ & $192 \times 1024 \times 512$ & $384 \times 2048 \times 1024$\\
       5   & $10^6$ - $10^8$ & $192 \times 1024 \times 512$ & $576 \times 3072 \times 1536$\\
       10  & $10^6$ - $10^9$ & $256 \times 1024 \times 512$ & $768 \times 3072 \times 1536$\\
       100 & $10^7$ - $10^9$ & $256 \times 1024 \times 512$ & $768 \times 3072 \times 1536$\\
  \end{tabular}
  \caption{
      Overview of the dimensionless parameters and grid resolutions used in the numerical simulations.
      Grid resolutions are listed here for the cases at highest $\Ray$, and we distinguish between the base grid used to evolve the velocity and the refined grid used to evolve the temperature field.
    }
  \label{tab:sims}
  \end{center}
\end{table}

The governing equations \eqref{eq:NS1}-\eqref{eq:NS2} are solved numerically using a second-order finite difference scheme for spatial derivatives and a third-order Runge--Kutta scheme for time stepping, as described in \cite{verzicco_finite-difference_1996} and \cite{van_der_poel_pencil_2015}.
For high values of $\Pran$, the temperature field must be resolved at smaller scales than the velocity field because the temperature field diffuses on the timescale of the order of $\Pran^{-1}$ compared to the velocity field.
We therefore also use the multiple-resolution technique of \cite{ostilla-monico_multiple-resolution_2015} to evolve the scalar $T$ on a refined grid.
Interpolation between the two grids is achieved through a four-point Hermitian method.
Grid stretching is also implemented in the wall-normal ($x$) direction using a clipped Chebyshev-type clustering.
Uniform grid spacing is used in the $y$ and $z$ directions, and the base grid of all simulations are resolved down to a factor of 2 times the Kolmogorov scale.
The refined grid is such that the wall-normal grid spacing satisfies $\Delta_x <0.5L_B$ at the boundaries, and the grid spacing in the bulk satisfies $\Delta_{x,y,z} < 4.5L_B$, where $L_B = (\nu\kappa^2/\eps)^{1/4}$ is the Batchelor scale.

The range of dimensionless control parameters simulated is shown in table \ref{tab:sims}.
Simulations at $\Ray=10^6$ are initialised using the laminar, purely conductive solution of \citet{batchelor_heat_1954} with the addition of small amplitude random noise to trigger a transition to turbulence.
Simulations at higher $\Ray$ are initialised using the final state of the simulation at $\Ray=10^6$ and $\Pran=1$, interpolated onto a new grid.
Each computation is performed for at least $300$ free-fall times, where $H/U_T$ is the free-fall time unit.
We average statistics over the last $250$ time units once the system has reached a statistically steady state.

\edit{
  \subsection{Response parameters and theoretical scaling laws}
  Before presenting the results of the simulations, we now provide an overview of the key quantities of interest and existing theoretical frameworks used for their prediction.
}

Understanding how the global horizontal heat transport in vertical convection depends on the control parameters \edit{of \eqref{eq:control_def}} is vital for many applications.
Varying the control parameters also leads to changes in the peak velocity of the rising flow and the mean shear stress on the boundary.
These can be quantified through the following dimensionless response parameters: the Nusselt number, the Reynolds number, and the shear Reynolds number
\begin{align}
    \Nuss &= \frac{H q_T}{\kappa \Delta}, &
    \Rey &= \frac{V_\mathrm{max} H}{\nu} , &
    \Rey_\tau &= \frac{V_\ast H}{\nu} , \label{eq:response}
\end{align}
where $q_T=\kappa\left|d\overline{T}/dx\right|_\mathrm{wall}$ is the horizontal heat flux, $V_\mathrm{max}$ is the peak value of the time- and spatially-averaged vertical velocity $\overline{v}(x)$, and $V_\ast = \sqrt{\tau_w/\rho_0}$ is the friction velocity associated with the mean wall shear stress $\tau_w=\mu \left. d\overline{v}/dx\right|_\mathrm{wall}=\rho_0 {V_\ast}^2$.

\edit{
  In turbulent convection, many studies follow the so-called `classical' regime as a theoretical starting point.
  This regime relies on the assumption that the thermal driving is sufficiently strong such that the heat flux becomes independent of the plate separation $H$.
  Assuming a power-law relation between $\Nuss$ and the Rayleigh number, dimensional analysis \citep[e.g.][]{turner_buoyancy_1979} then requires the scaling $\Nuss \sim \Ray^{1/3}f(\Pran)$.
  This has been consistent with various experiments up to $Ra = 10^{12}$ \citep{warner_experimental_1968,tsuji_characteristics_1988} and is often provided in engineering reference texts such as \citet{holman_heat_2010}.
}

\edit{
  However, recent analysis of numerical simulations by \citet{ng_changes_2017} suggests that a power-law description may be insufficient and that the `classical' scaling does not accurately describe the data even in this range.
  Furthermore, there are open questions regarding the relevant scaling at even higher $\Ray$, at which precise, controlled experiments and numerical simulations are extremely difficult to perform.
  Finally, the Prandtl number dependence has hardly been addressed.
}

One important application for boundary layers in vertical convection is to predict the dissolution or melting of a vertical ice face in the ocean.
This is why parameterising the heat and salt fluxes is crucial.
In regional ocean models, the heat flux through the turbulent boundary layer at such locations is often parameterised by invoking the heat flux balance $q_T \sim \textrm{velocity} \times \textrm{temperature change}$.
Following \citet{holland_modeling_1999}, the parameterisation for the horizontal heat flux takes the form
\begin{equation}
  q_T = C_T C_D^{1/2} U  (T - T_b), \label{eq:heat_flux}
\end{equation}
where $U$ is the vertical velocity of the rising plume, and $T-T_b$ is the temperature difference between the ocean and the ice boundary.

Taking $U=V_\mathrm{max}$ and $T-T_b = \Delta/2$, we note that the drag coefficient $C_D$ and `transfer coefficient' $C_T$ from \eqref{eq:heat_flux} are fully determined by the response parameters of \eqref{eq:response} through
\begin{align}
    C_D &= \left(\frac{V_\ast}{V_\mathrm{max}}\right)^2 = \frac{{\Rey_\tau}^2}{{\Rey}^2} , &
    C_T &= \frac{2q_T}{V_\ast \Delta} = \frac{2\Nuss}{\Rey_\tau \Pran}. \label{eq:CdGamma}
\end{align}
Accurate scaling laws for the quantities in \eqref{eq:response} are thus crucial for determining $C_D$ and $C_T$.
The transfer coefficient $C_T$ is equivalent to a modified Stanton number where $V_\ast$ is used for the velocity scale.
In the ice-ocean literature, the transfer coefficient is often denoted $\Gamma_T$ although we  use $C_T$ here to avoid confusion with the aspect ratio $\Gamma$ used throughout literature on convection.
Both $C_D$ and $C_T$ are typically set to constant values in melt parameterisations \citep[see e.g.][]{jackson_meltwater_2020} based on the reasoning that the boundary layers in ice-ocean applications are strongly shear-driven, and are in accordance with the classical results of \citet{kader_heat_1972}.
However recent analysis by \citet{malyarenko_synthesis_2020} of ice shelf observations suggests that the Reynolds numbers may not always be large enough to justify this shear-driven boundary layer assumption.
An equivalent equation to \eqref{eq:heat_flux} is often used to parameterise the salt flux, where $C_D$ keeps the same value, but $C_T$ is reduced to reflect its dependence on the Schmidt number.

\edit{
  For $C_T$, $C_D$ being constant, the dimensionless form of \eqref{eq:heat_flux} is $\Nuss\sim\Rey\Pran$.
  Such a scaling is reminiscent of the `ultimate' or `diffusion-free' scaling hypothesised for Rayleigh-B\'enard convection (RBC) at very high $\Ray$ \citep[e.g.][]{kraichnan_turbulent_1962,spiegel_convection_1971,lohse_ultimate_2003,ahlers_heat_2009}.
  In that case, the heat flux is assumed independent of the molecular diffusivities $\nu$ and $\kappa$, such that dimensional analysis implies $\Nuss \sim (RaPr)^{1/2}$.
  In physical terms, this regime is associated with a dominant large-scale circulation that leads to shear-driven turbulent boundary layers.
  The dominant mean flow arising in VC is analogous to such a coherent large-scale circulation \citep{shishkina_thermal_2016}.
  RBC provides a useful comparison to VC thanks to its identical geometry (except for the direction of gravity) and its dependence on the same control parameters.
}

\edit{
  In the case of RBC, a unifying theory describing the transitions between various regimes in RBC was proposed by \citet{grossmann_scaling_2000,grossmann_thermal_2001}.
  This theory has shown excellent agreement with subsequent experimental and numerical investigations over a large range of $\Ray$ and $\Pran$ \citep{ahlers_heat_2009,stevens_unifying_2013}.
  Although VC lacks the global relation between the Nusselt number and the mean dissipation rate of kinetic energy required to close the equations corresponding to those of the Grossmann-Lohse (GL) theory, it remains appealing to search for parallels between RBC and VC to understand how the heat flux can be parameterised as the boundary layers evolve.
  \citet{wells_geophysical-scale_2008} applied ideas from the GL theory about boundary layer transition to geophysical-scale convection at a vertical wall, and \citet{ng_vertical_2015,ng_changes_2017} considered how changes in the boundary layer structure relate to an increased bulk contribution to turbulent dissipation.
  However, these studies left the issue of $\Pran$-dependence largely unresolved.
  In this study, we aim to gain insight on how the Prandtl number affects (a) the scaling of the above response parameters in the currently accessible range of $\Ray$, and (b) any subsequent transition in the nature of the boundary layers.
}

\section{Flow visualisation} \label{sec:viz}

\begin{figure}
  \centerline{\includegraphics[width=\linewidth]{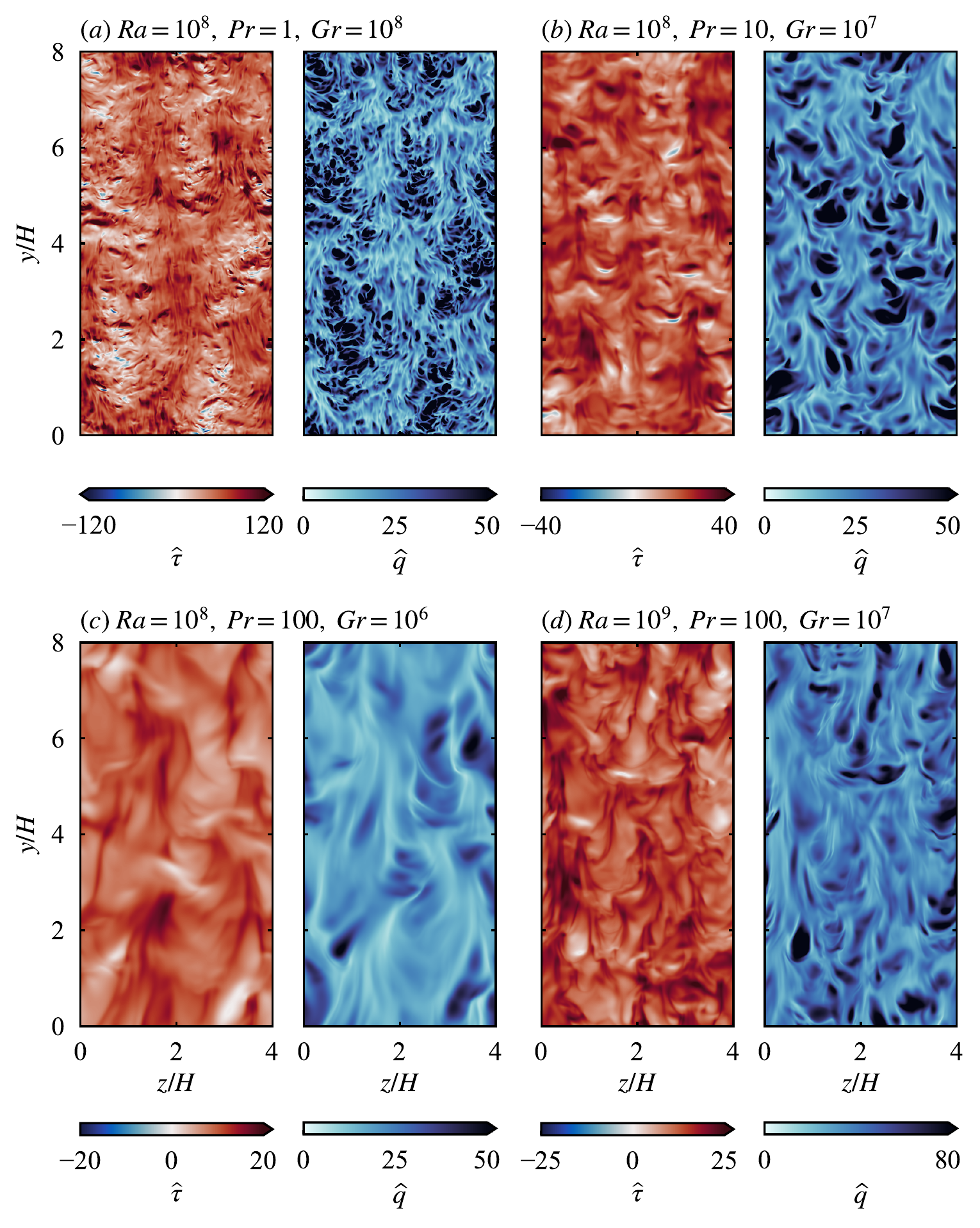}}% Images in 100% size
  \caption{
    Final-time snapshots of the dimensionless horizontal shear stress $\widehat{\tau}$ and the local dimensionless heat flux $\widehat{q}$ at the heated wall $x=0$ for a range of $\Ray$ and $\Pran$.
    It can be seen how large $\Pran$ smoothes the fields, even at a large $\Ray=10^8$.
  }
\label{fig:yz_planes}
\end{figure}

To illustrate how the boundary layers in the flow change with $\Pran$ and $\Ray$, we present a snapshot of the local dimensionless vertical shear stress $\widehat{\tau}$ and heat flux $\widehat{q}$ at the heated wall $x=0$ in figure \ref{fig:yz_planes}.
These quantities are defined as
\begin{align}
    \widehat{\tau}(y,z) &= \frac{H}{U_T} \left.\frac{\pd v}{\pd x}\right|_{x=0}, &
  \widehat{q}(y,z) &= -\frac{H}{\Delta} \left.\frac{\pd T}{\pd x}\right|_{x=0}.
\end{align}
We note that averaging $\widehat{q}$ over the plane and over time gives the Nusselt number $\langle \widehat{q}(y,z,t)\rangle_{y,z,t}=\Nuss$.
In this sense, $\widehat{q}$ can be thought of as a `local and instantaneous Nusselt number'.
The snapshots, taken at the end time of each simulation, highlight the striking localisation of the heat flux at the wall.
Consistent with the analysis of \citet{pallares_turbulent_2010}, the regions of strongest heat flux (being the dark patches in the panels of $\widehat{q}$) are frequently co-located with instantaneous flow reversals (evidenced by white and blue patches appearing in the panels for $\widehat{\tau}$).

Panels ($a$)-($c$) of figure \ref{fig:yz_planes} highlight the effect of increasing $\Pran$ in this setup while keeping $\Ray$ fixed (in this case at $10^8$).
As seen from the colour scales, although the range of the local Nusselt number $\widehat{q}$ remains similar as $\Pran$ increases, a significant decrease in the mean dimensionless shear stress is observed at high $\Pran$.
This is due to a drop in the Grashof number $\Gr$ as $\Pran$ is increased for fixed $\Ray$.
The Grashof number quantifies the ratio of buoyancy effects to viscosity, and is analogous to a squared Reynolds number based on the free-fall velocity.

This analogy with the Reynolds number provides some further intuition for the snapshots of figure \ref{fig:yz_planes}, where a much wider range of length scales can be observed in the $\widehat{\tau}$ field for the high $\Gr$ snapshot of ($a$) compared to the lower $\Gr$ snapshot of ($c$).
By contrast, comparing panels ($b$) and ($d$) allows us to visualise the effect of changing $\Pran$ while keeping $\Gr$ fixed.
Qualitatively the structures in both the $\widehat{\tau}$ and $\widehat{q}$ snapshots appear similar.
However, the mean values of both quantities vary as $\Pran$ increases.
\edit{
  To obtain a more quantitative evaluation of the boundary layer structures and to more quantitatively extract length scales, we have also calculated the relevant power spectra for each of the simulations shown in figure \ref{fig:yz_planes}.
  These results (not shown here) emphasise the similarity of structures for constant $\Gr$ at the walls, although this similarity does not extend outside of the viscous boundary layer.
}

Compared to Rayleigh-B\'enard convection (RBC), where large-scale thermal structures do not exhibit a preferred direction, the mean shear at the wall in VC introduces significant anisotropy to the wall structures.
Streaky structures elongated in the vertical ($y$) direction are prominent in figure \ref{fig:yz_planes}, similar to those seen in the sheared RBC setup of \citet{blass_effect_2021}.
Furthermore, in that study, an increased $\Pran$ (for fixed $\Ray$ and $\Rey$) was found to enhance momentum transport from the walls, allowing the wall shear to affect the flow structures in the bulk more easily.
However, as mentioned in \S{2}, the wall shear in VC is not pre-determined and instead arises as a response parameter of the system.

The snapshots of figure \ref{fig:yz_planes} highlight the complex multi-parameter dependence in the vertical convection setup.
Indeed, the simple analogy between the Grashof number and the square of the Reynolds number should not be overstated.
As shown in figure \ref{fig:mean_profs}($b$), the peak value of the time-averaged vertical velocity does not simply scale with the free-fall velocity $U_T$, but varies depending on $\Pran$.
In the following section, we shall investigate the multi-parameter dependence more quantitatively by identifying scaling relations for key response parameters of the system.
% \citet{blass_effect_2021}

\section{Heat flux and Reynolds number parameterisation} \label{sec:heat_flux}

\begin{table}
  \begin{center}
\def~{\hphantom{0}}
  \begin{tabular}{cccc}
    Response parameters               & Two parameter regression                                       & \citet{shishkina_momentum_2016} & \GL\ $\mathrm{IV}_u$ \\[3pt]
    Nusselt number $\Nuss$            & $\Ray^{\edit{ 0.321 \pm 0.006}} \Pran^{\edit{-0.083 \pm 0.010}}$ & $\Ray^{ 1/4}$              & $\Ray^{ 1/3}$ \\[3pt]
    Reynolds number $\Rey$            & $\Ray^{\edit{ 0.489 \pm 0.007}} \Pran^{\edit{-0.738 \pm 0.010}}$ & $\Ray^{ 1/2} \Pran^{-1}$   & $\Ray^{ 4/9} \Pran^{-2/3}$ \\[3pt]
    Shear Reynolds number $\Rey_\tau$ & $\Ray^{\edit{ 0.362 \pm 0.002}} \Pran^{\edit{-0.446 \pm 0.003}}$ & $\Ray^{ 3/8} \Pran^{-1/2}$ & $\Ray^{ 1/3} \Pran^{-1/2}$ \\[3pt]
    Drag coefficient $C_D$            & $\Ray^{\edit{-0.253 \pm 0.010}} \Pran^{\edit{ 0.584 \pm 0.015}}$ & $\Ray^{-1/4} \Pran$        & $\Ray^{-2/9} \Pran^{ 1/3}$ \\[3pt]
    Transfer coefficient $C_T$        & $\Ray^{\edit{-0.041 \pm 0.006}} \Pran^{\edit{-0.637 \pm 0.009}}$ & $\Ray^{-1/8} \Pran^{-1/2}$ &             $\Pran^{-1/2}$
  \end{tabular}
  \caption{
      Observed effective scalings laws for various dimensionless response parameters.
      Only simulations with $\Rey>150$ are included in the linear regression.
      \edit{The uncertainty shown is the standard deviation of the estimated slopes, as described in the text of \S4.}
      Theoretical scaling relations for laminar VC and turbulent RBC from \citet{shishkina_momentum_2016} \edit{for VC} and \citet{grossmann_scaling_2000} \edit{for RBC in the so-called $\mathrm{IV}_u$} are provided for comparison.
      $\Rey_\tau$ is calculated for these scaling relations using the similarity variable of \citet{shishkina_momentum_2016} and using the Blasius drag law $C_D\sim Re^{-1/2}$ for the GL theory.
    }
  \label{tab:scalings}
  \end{center}
\end{table}

\begin{figure}
  \centerline{\includegraphics[width=\linewidth]{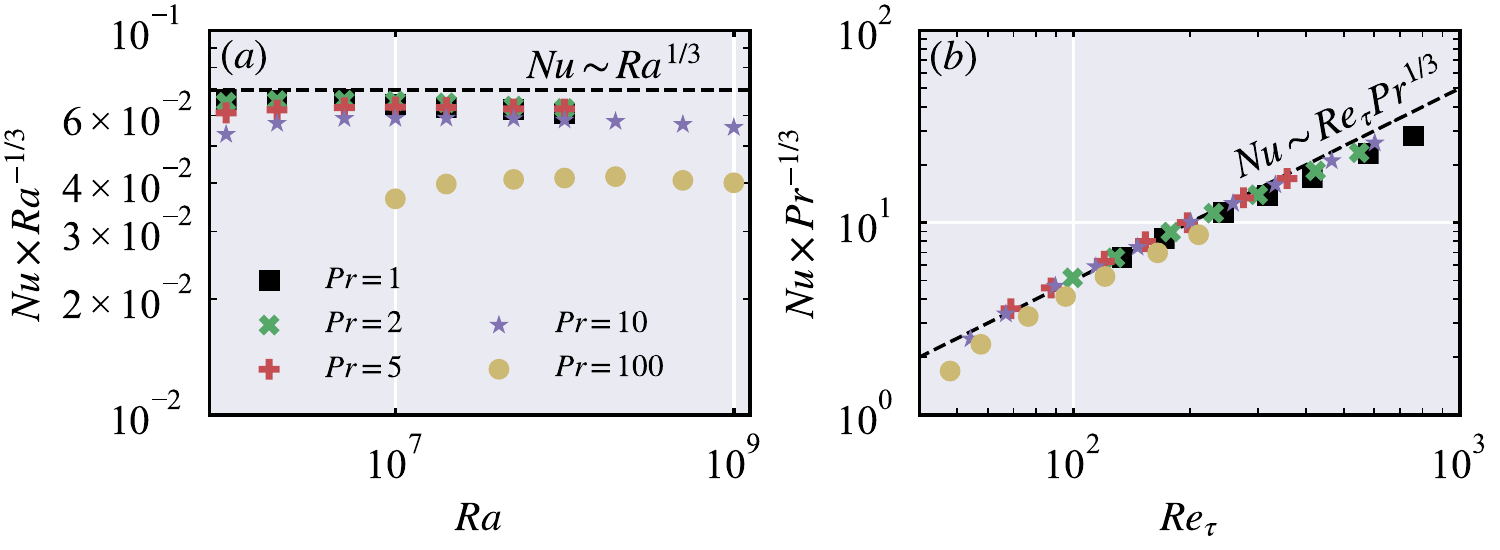}}
  \caption{
    Nusselt number against $(a)$ Rayleigh number (compensated by $\Ray^{1/3}$), and $(b)$ against shear Reynolds number (compensated by $\Pran^{1/3}$).
    % Error bars denote standard deviations in the time series of $\langle u^\prime T^\prime\rangle$.
    % The black dashed line represents the theoretical scaling of $Nu\sim Ra^{1/3}$, which panel $(b)$ is compensated by.
  }
\label{fig:NuRa}
\end{figure}

In table \ref{tab:scalings} we report the observed $\Ray$- and $\Pran$-dependence of the response parameters from \eqref{eq:response} and \eqref{eq:CdGamma} in our simulations.
An effective power-law dependence is assumed and two-parameter linear regression is used to obtain the effective scaling exponents.
\edit{
  Precisely, we compute $\mathbf{b} = X^{-1}\mathbf{y}$, where $\mathbf{b}=(b_1, b_2, b_3)^T$ and
  \begin{align}
    x_{i 1} &= \log \Ray_i, &
    x_{i 2} &= \log \Pran_i, &
    x_{i 3} &= 1, &
    y_i &= \log R_i, &
    i=1,\dots,n
  \end{align}
  are constructed from the $n$ simulations for each response parameter $R$, giving a linear fit $R=\Ray^{b_1} \Pran^{b_2} 10^{b_3}$.
  We calculate the uncertainty of the power law exponents $b_1$ and $b_2$ through the variance matrix of $\mathbf{b}$ given by $V=\sigma^2 (X^T X)^{-1}$, where $\sigma^2$ is the variance of $\mathbf{y} - X \mathbf{b}$.
  The standard deviations of the slopes, given by $\sqrt{v_{11}}$ and $\sqrt{v_{22}}$ are presented in table \ref{tab:scalings}.
}

The Nusselt number is consistent with the theoretical scaling relation $\Nuss\sim\Ray^{1/3}f(\Pran)$ that arises when the heat flux is assumed to be independent of the plate separation \citep{malkus_heat_1954}.
\citet{ng_changes_2017} suggested that for $\Pran\approx 1$, a regime transition to a shear-dominated boundary layer is underway at $\Ray=10^9$, but following \citet{grossmann_scaling_2000}, this transitional $\Ray$ can be expected to increase with $\Pran$, as the smaller Reynolds number stabilizes the flow.
Our results contrast with the effective scaling laws for laminar vertical convection derived by \citet{shishkina_momentum_2016}, where $\Nuss\sim\Ray^{1/4}$ and $\Rey\sim\Ray^{1/2}\Pran^{-1}$ for $\Pran\gg 1$.
This difference is to be expected since our setup is far from the laminar state for which the scaling laws have been observed to hold \citep[e.g. by][]{wang_regime_2021}.

In figure \ref{fig:NuRa} we plot $\Nuss$ against both $\Ray$ and the shear Reynolds number $\Rey_\tau$.
Figure \ref{fig:NuRa}a highlights the weak dependence of $\Nuss$ on $\Pran$, with higher $\Pran$ typically reducing $\Nuss$ for a fixed value of $\Ray$.
\edit{
  Note that a simple, single power-law fit is unlikely to adequately describe the heat transfer outside of the currently accessible parameter range.
  Even within the data presented here, the $\Pran=1$ cases appear to trend downwards relative to the $\Ray^{1/3}$ line on figure \ref{fig:NuRa}a at higher values of $\Ray$.
  This observation is consistent with \citet{ng_changes_2017}, who attribute the decrease to a lower heat flux contribution from regions of weak shear.
  Later in this section, and in appendix \ref{app:BL_trans}, we shall discuss at which parameter values we may expect a transition to shear-driven turbulent boundary layers and how this would affect the scaling of the Nusselt number.
}

Against $\Rey_\tau$ in figure \ref{fig:NuRa}b, we obtain a reasonable collapse for $\Nuss$ by scaling with $\Pran^{1/3}$ and observe a scaling close to $\Nuss\sim\Rey_\tau \Pran^{1/3}$.
Since this is consistent with the high $\Pran$ limit of passive heat transport in turbulent boundary layers from \citet{kader_heat_1972}, we are motivated to compare with passive scalar transport in other turbulent flows.
For example, a recent study by \cite{yerragolam_passive_2021} proposed a scaling theory for passive scalar transport in plane Couette flow where $\Nuss\sim\Rey_\tau^{6/7}\Pran^{1/2}$.
%$\Rey_\tau\sim \Rey^{7/8}$ and $\Nuss\sim\Pran^{1/2}\Rey^{3/4}$.
%In that system, the Reynolds number is a control parameter determined by the velocity of the walls.
% Combining the two scaling relations from \citet{yerragolam_passive_2021} gives .
This somewhat contrasts with the $\Pran^{1/3}$ collapse observed in figure \ref{fig:NuRa}b, although the higher $\Rey_\tau$ values of our data do exhibit a local scaling exponent less than one and close to $6/7$.
% Further research is needed to understand whether, at higher $\Ray$, an analogy can be drawn between VC and Couette flow for $\Nuss(\Rey_\tau,\Pran)$.

\begin{figure}
  \centerline{\includegraphics[width=\linewidth]{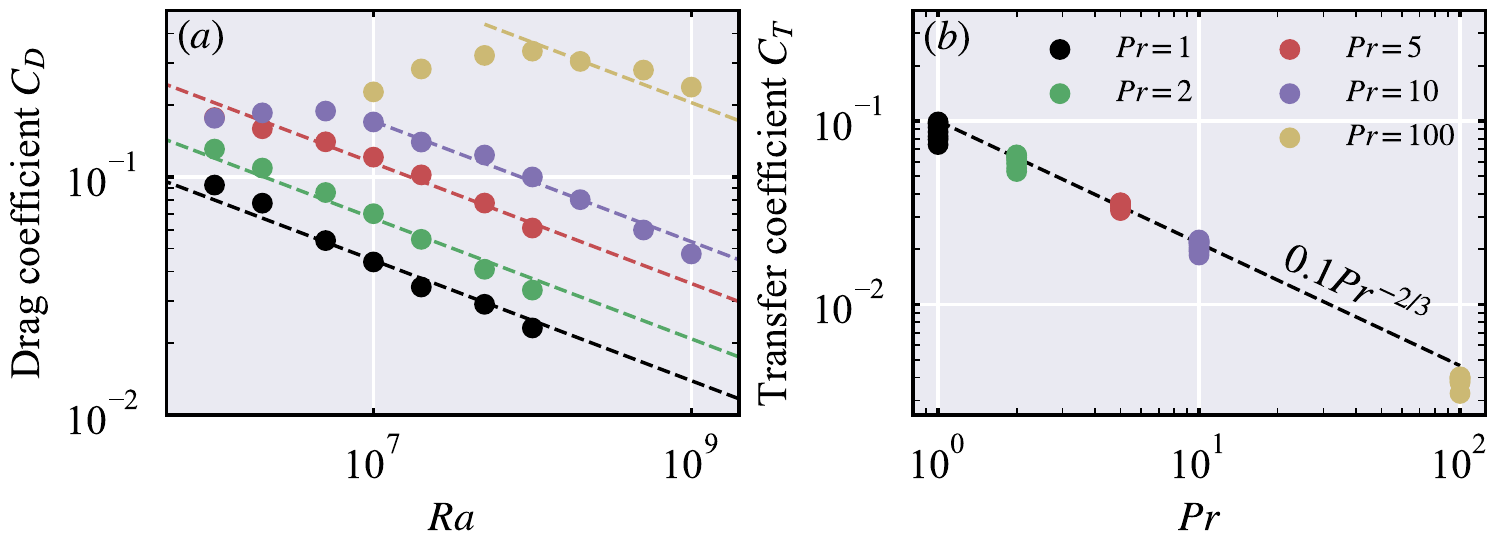}}
  \caption{
    Drag coefficient $C_D$ and transfer coefficient $C_T$ defined in \eqref{eq:CdGamma}.
    Dashed lines in panel $(a)$ use the exponents obtained from the linear regression in table \ref{tab:scalings}.
  }
\label{fig:melt_params}
\end{figure}

We note that the Reynolds number scaling in table \ref{tab:scalings} is close to that reported by \citet{lam_prandtl_2002} from experiments of Rayleigh-B\'enard convection with a range of large Prandtl numbers.
\citet{lam_prandtl_2002} suggested that their results were consistent with the theoretical scaling relation $\Rey\sim \Ray^{4/9}\Pran^{-2/3}$ proposed for the regime ($IV_u$) associated with ${\Nuss\sim\Ray^{1/3}}$ in the `GL theory' of \citet{grossmann_scaling_2000,grossmann_thermal_2001}, although \citet{lam_prandtl_2002} acknowledge that this measured effective $\Pran$ exponent shows a relatively large deviation from the theory.
Furthermore, these deviations varied depending on the definition of the Reynolds number inferred from their experiments.
We note that the $\Rey\sim\Ray^{4/9}\Pran^{-2/3}$ scaling can also be derived from dimensional analysis by assuming that the vertical velocity $V_\mathrm{max}$ is solely determined by the buoyancy flux per unit area $\Phi=g\alpha q_T$ and the plate separation $H$ \citep[as in the `outer' scaling of][]{george_theory_1979}, and also assuming the \citet{malkus_heat_1954} scaling $\Nuss\sim\Ray^{1/3}$.
As seen from table \ref{tab:scalings}, this $\Rey$ scaling does not perfectly capture the observed data, and we cannot rule out the effect of multiple regimes on the effective scaling exponent, like in the GL theory for RBC.
More work is needed to provide a theoretical understanding for these results.

As highlighted by \citet{mcconnochie_testing_2017}, the scaling relation $\Nuss \sim \Ray^{1/3}f(\Pran)$ implies \edit{a dimensional form for the heat flux that scales as $F_T\sim \Delta T^{4/3}$ for fixed fluid properties.
T}he heat flux is \edit{therefore} independent of the bulk velocity $V_\mathrm{max}$, making \edit{the shear-based model of} \eqref{eq:heat_flux} an inappropriate parameterisation for this regime.
Indeed, as shown in figure \ref{fig:melt_params}, we observe significant variation in the drag coefficient $C_D$ with both $\Ray$ and $\Pran$.
In all cases we find a value much larger than the high-$\Rey$ limit of $C_D=2.5\times 10^{-3}$, as used by \citet{holland_modeling_1999}.
However, the scaling observed for the transfer coefficient $C_T \approx 0.1 \Pran^{-2/3}$ \emph{is} consistent with the values used for parameterising heat and salt fluxes in that work and subsequent melting studies.
Using the definition from e.g.\ \eqref{eq:CdGamma}, we can express this result in terms of the Nusselt number as $\Nuss \sim \Rey_\tau \Pran^{1/3}$ or with dimensional quantities as $q_T \sim \Pran^{-2/3} V_\ast \Delta$.

\begin{figure}
  \centerline{\includegraphics[width=\linewidth]{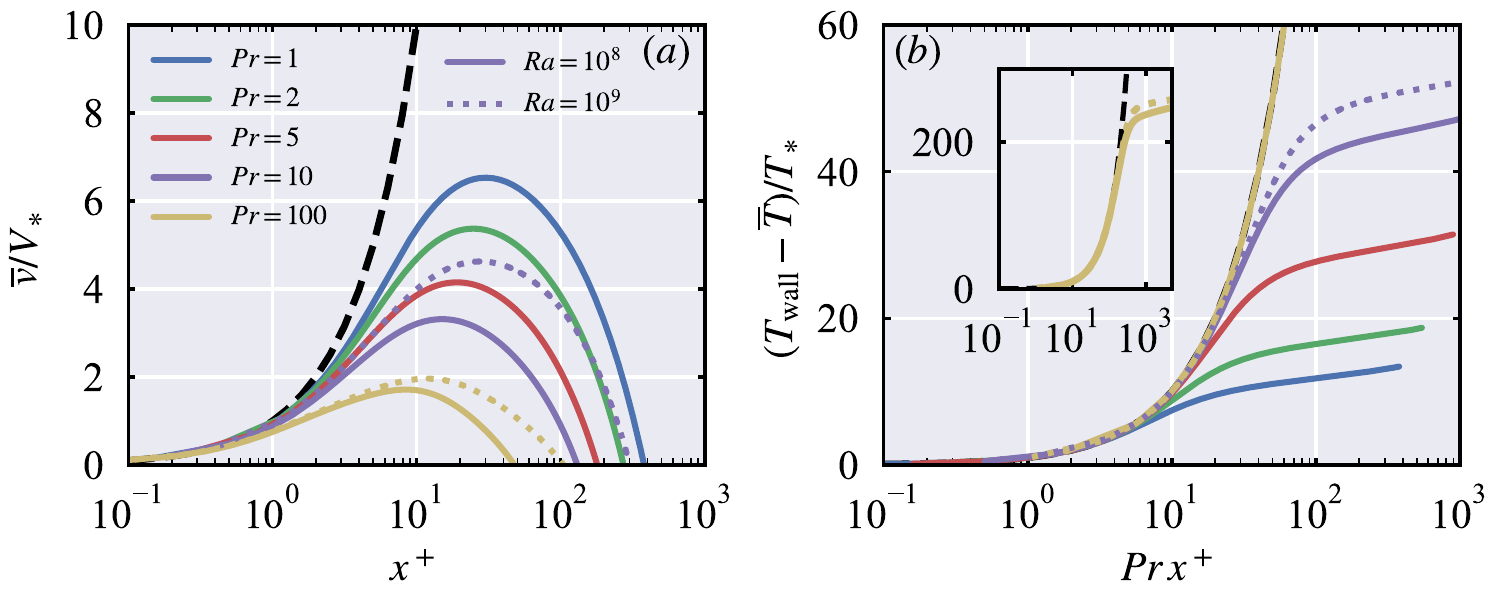}}
  \caption{
    \edit{
      Mean profiles of $(a)$ vertical velocity and $(b)$ temperature on a logarithmic $x$ axis.
      The $x$-axis is scaled in terms of viscous wall units, such that $x^+ = xV_\ast/\nu$.
      Vertical velocity is scaled by the friction velocity $V_\ast$, and temperature is scaled by the equivalent `friction temperature' scale $T_\ast = q_T/V_\ast$.
      Solid lines denote simulations at $\Ray=10^8$, whereas dotted lines represent the two simulations at $\Ray=10^9$.
      The dashed black lines denote the linear profiles $\overline{v}=V_\ast x^+$ and $\overline{T} = T_\ast \Pran x^+$ in panels $(a)$ and $(b)$ respectively.
      The inset in $(b)$ is a zoom-out of the main figure highlighting the results for $\Pran=100$.
    }
  }
  \label{fig:log_BLs}
\end{figure}

\edit{
  It may be tempting to associate the scaling $\Nuss\sim\Rey_\tau \Pran^{1/3}$ with the appearance of turbulent boundary layers in the sense of Prandtl and von K\`arm\`an, where log-law profiles appear in the mean velocity and temperature profiles.
  However, this is not the case for our simulations.
  In figure \ref{fig:log_BLs} we plot these mean profiles from the simulations at $\Ray=10^8, 10^9$ with a logarithmic $x$-axis.
  From figure 5a, it is clear that log-layers are absent from the velocity profile.
  Indeed, we are far from the critical Reynolds number for transition to such a shear-driven boundary layer.
  As we explore in appendix \ref{app:BL_trans}, Rayleigh numbers above $10^{11}$ are likely to be necessary for this transition and such critical values only increase with $\Pran$.
  By contrast, the temperature profiles of figure 5($b$) appear consistent with logarithmic profiles.
  This observation is somewhat unsurprising, given the appearance of such profiles in the `classical' regime of RBC by \citet{ahlers_logarithmic_2012}.
  A logarithmic profile in the temperature field does \emph{not} imply the presence of a shear-driven turbulent boundary layer.
}

\citet{holland_modeling_1999} associate the scaling relation $C_T \sim Pr^{-2/3}$ with the strong influence of a molecular sublayer where conduction is the dominant mechanism of heat transport.
Motivated by this result, we proceed by investigating how the width of this boundary layer depends on the control parameters of the vertical convection system.

\section{Conductive thermal boundary layer} \label{sec:BLs}

In a statistically steady state, the mean velocity and temperature profiles of the system satisfy
\begin{align}
    \frac{\d}{\d x} \overline{u^\prime \bu^\prime} &= \nu \frac{\d^2 \overline{\bu}}{\d x^2} + \overline{T} \hat{\mathbf{y}} , &
    \frac{\d}{\d x} \overline{u^\prime T^\prime} &= \kappa \frac{\d^2 \overline{T}}{\d x^2} , \label{eq:T_mean}
\end{align}
where an overbar denotes an average in $y$, $z$, and time.
Incompressibility ensures that $\overline{u}\equiv 0$, so the mean velocity $\overline{\bu}$ only has components in the wall-parallel directions.
The second equation of \eqref{eq:T_mean} implies that the heat flux at any wall-normal location must be constant, or in dimensionless terms
\begin{equation}
    \Nuss = \frac{H}{\kappa\Delta} \left(\overline{u^\prime T^\prime} - \kappa \frac{\d\overline{T}}{\d x}\right) = \mathrm{constant}. \label{eq:Nusselt}
\end{equation}
Following \cite{wells_geophysical-scale_2008} and in the spirit of \citet{grossmann_scaling_2000}, we divide the flow into thermal boundary layers, where the heat flux is dominated by molecular diffusion of the mean, and bulk regions, where the heat flux is due to the `wind' of turbulence.
Precisely, we define the conductive thermal boundary layer width $\delta_T$ as the wall-normal location where the conductive heat flux $-\kappa d\overline{T}/dx$ is equal to the turbulent heat flux $\overline{u^\prime T^\prime}$.

In Rayleigh-B\'enard convection at moderate $\Ray$, there is a general consensus from existing literature \citep{ahlers_heat_2009,ching_velocity_2019} that scaling-wise the thickness of the boundary layers follows a laminar-like scaling according to Prandtl, Blasius and Pohlhausen, that is
\begin{equation}
    \frac{\delta_T}{H} \sim \Rey^{-1/2} f(\Pran) . \label{eq:Blasius_BL}
\end{equation}
For vertical convection, \cite{ng_vertical_2015} suggested the application of the same form as \eqref{eq:Blasius_BL} at moderate Reynolds number, although only cited `fair' agreement with their DNS reporting an effective $\Rey$-exponent of $-0.60$.
The scaling \eqref{eq:Blasius_BL} is applicable in the case of a fully laminar flow as studied by \citet{kuiken_asymptotic_1968}, who derived an equivalent scaling of $\delta_T/H \sim \Gr^{-1/4}$ in the limit of high $\Pran$.%, where $\Gr=\Ray/\Pran$ is the Grashof number.
The scaling laws $\Nuss\sim\Ray^{1/4}$, $\Rey\sim\Ray^{1/2}Pr^{-1}$ proposed by \citet{shishkina_momentum_2016} are also consistent with \eqref{eq:Blasius_BL}.

\begin{figure}
  \centerline{\includegraphics[width=\linewidth]{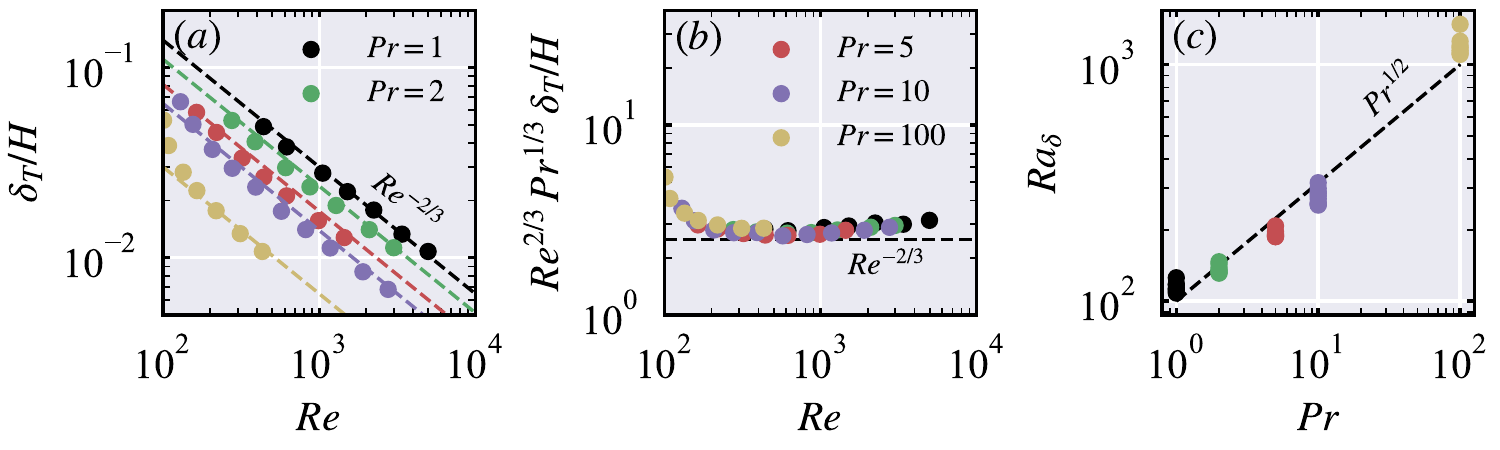}}
  \caption{
    $(a)$ Dimensionless conductive thermal boundary layer width $\delta_T/H$ against Reynolds number.
    $(b)$ Plot of the same data compensated by $\Rey^{2/3} \Pran^{1/3}$.
    $(c)$ Measured sublayer Rayleigh number $\Ray_\delta$ as a function of Prandtl number.
    Dashed lines in panel $(a)$ mark the suggested $\Rey^{-2/3} \Pran^{-1/3}$ scaling.
  }
\label{fig:T_BLs}
\end{figure}

From our new simulations, we find a collapse of the data such that $\delta_T/H \sim \Pran^{-1/3} f(\Rey)$, as shown in figure \ref{fig:T_BLs}.
This $\Pran$-dependence is well known from the similarity scaling of a laminar boundary layer at a horizontal wall \citep[e.g.][]{schlichting_boundary-layer_2016}, applied to the regimes of \citet{grossmann_scaling_2000} where the thermal dissipation rate is dominated by boundary layer contributions.
However the $\Pran^{-1/3}$ factor does \emph{not} arise in the laminar solutions for VC considered by \citet{kuiken_asymptotic_1968} and \citet{shishkina_momentum_2016}.
The $\Pran^{-1/3}$ scaling is often also observed in empirical data for turbulent flows \citep[e.g.][]{kader_temperature_1981}.
Indeed, rather than observing a laminar-like $\Rey^{-1/2}$ scaling, our data is more consistent with
\begin{equation}
    \frac{\delta_T}{H} \sim \Rey^{-2/3} \Pran^{-1/3} , \label{eq:TBL_scale}
\end{equation}
as shown in figure \ref{fig:T_BLs}($b$).

For the case where $V_\mathrm{max}\sim U_T$, the scaling of \eqref{eq:TBL_scale} is equivalent to $\delta_T/H \sim \Ray^{-1/3}$ and one can interpret the boundary layer width as being set by a critical Rayleigh number.
This is the `buoyancy-controlled sublayer' scaling as described by \citet{wells_geophysical-scale_2008}, similar to the marginally stable boundary layer argument of \citet{malkus_heat_1954} for Rayleigh-B\'enard convection.
However, as we already mentioned earlier, $V_\mathrm{max}$ does not simply scale with $U_T$ in our simulations.
In figure \ref{fig:T_BLs}($c$), we plot the `sublayer Rayleigh number' $\Ray_\delta = g\alpha \delta_T^3 \Delta /\nu\kappa$, and find that this value is \emph{not} constant, but instead strongly depends on the Prandtl number.
Further studies are certainly needed to understand how to interpret these results.
It remains an open question whether a $\Pran$-dependent critical Rayleigh number is appropriate for limiting the conductive boundary layer width or whether the Reynolds number plays a more significant role.
% generic \cs{(maybe use the phrase: understand the universality of the scaling of \ldots)} the scaling of \eqref{eq:TBL_scale} is.
The addition of a spanwise mean flow to the system, forming a three-dimensional mixed convection setup, would allow $\Rey$ and $\Ray$ to be decoupled, and reveal the inherent parameter dependence of the boundary layer.

\section{Conclusions} \label{sec:discussion}

Through three-dimensional direct numerical simulations, we have investigated the multi-parameter dependence of convection in a vertical channel for Prandtl $1\leq \Pran \leq 100$ and Rayleigh numbers $10^6 \leq \Ray \leq 10^9$.
We observe Nusselt numbers consistent with the classical $\Ray^{1/3}$ scaling combined with some weak but non-trivial dependence on the Prandtl number.
The Reynolds number associated with the large scale `wind' exhibits a scaling of $\Ray^{0.491} \Pran^{-0.735}$, similar to that measured by \citet{lam_prandtl_2002} in experiments of Rayleigh-B\'enard convection.
The discrepancy between the observed scaling and the theoretical prediction of $\Rey\sim\Ray^{4/9}\Pran^{-2/3}$ from \citet{grossmann_scaling_2000,grossmann_thermal_2001} for RBC however suggests there is more work to be done to build a theoretical understanding for the behaviour of turbulent VC.
We cannot rule out the possibility that our observations arise due to a mixed scaling with contributions from multiple flow regimes.% as hinted by \citet{ng_changes_2017}.}

As previously highlighted by \citet{mcconnochie_testing_2017}, such a scaling for $\Nuss$ is inconsistent with the commonly used heat flux parameterisation of \citet{holland_modeling_1999}.
Our simulations highlight that this discrepancy is due to a highly variable drag coefficient in vertical convection that depends on both of the control parameters $\Ray$ and $\Pran$.
\edit{
  The absence of logarithmic velocity profiles suggests that the lack of shear-driven turbulent boundary layers is to blame for the large variation in the drag coefficient.
  By considering the critical Reynolds number of \citet{landau_fluid_1987} in appendix \ref{app:BL_trans}, we infer that transition to such turbulent boundary layers will only occur for $\Ray > 4\times 10^{11} \times \Pran^{1.89}$.
  However, more work is needed to understand how this transition occurs, and whether local scaling exponents for $\Nuss$ become impacted by multiple regimes and logarithmic corrections, as is the case for RBC \citep{grossmann_multiple_2011} and convection from rough walls \citep{macdonald_heat_2019}.
}

\edit{In contrast to the variation in the drag coefficient}, the transfer coefficient (or modified Stanton number) satisfies $C_T\approx 0.1 \Pran^{-2/3}$, matching values used in ice-ocean parameterisations.
In other words, the friction velocity $V_\ast$ in this flow seems to adjust such that the heat flux scales as $q_T\sim V_\ast \Delta$ for each given value of $\Pran$.
The strong dependence of $C_T$ on $\Pran$ suggests that the conductive sublayer at the wall plays an important role in the total heat flux.
We diagnose the width of this sublayer from the simulations and find the scaling $\delta_T/H \sim {\Rey}^{-2/3}\Pran^{-1/3}$ to be consistent with our data.
The emergent Rayleigh number $\Ray_\delta$ associated with this sublayer is found to depend strongly on Prandtl number, questioning the notion of marginal stability at a critical value of $\Ray_\delta$.
This is similar to RBC, where the marginal stability theory of \citet{malkus_heat_1954} is also insufficient to fully describe the control parameter dependence of the heat flux \citep{ahlers_heat_2009}.

Understanding how generic these results are will be vital for environmental applications.
For example, \citet{jackson_meltwater_2020} recently highlighted the role of mean horizontal flows in enhancing heat and salt transport at melting ice faces.
In such a mixed convection scenario, $\Rey$ is not necessarily coupled to $\Ray$ as it is in vertical convection.
Thus understanding the underlying parameter dependence is an important topic for future research.
As reviewed by \citet{malyarenko_synthesis_2020}, many factors not considered here can also be important for the ablation of ice in the ocean.
In particular, the presence of both temperature and salinity variations and the dynamic melting condition may modify the nature of the boundary layers in this geophysical setting.

% \backsection[Supplementary data]{\label{SupMat}Supplementary material and movies are available at \\https://doi.org/10.1017/jfm.2019...}

% \backsection[Acknowledgements]{Acknowledgements may be included at the end of the paper, before the References section or any appendices. Several anonymous individuals are thanked for contributions to these instructions.}

\backsection[Funding]{
This project has received funding from the European Research Council (ERC) under the European Union’s Horizon 2020 research and innovation programme (Grant agreement No. 804283).
We acknowledge PRACE for awarding us access to Joliot-Curie at GENCI@CEA, France, and this work was also sponsored by NWO Science for the use of supercomputer facilities.
% Please provide details of the sources of financial support for all authors, including grant numbers. For example, "This work was supported by the National Science Foundation (grant number XXXXXXX)". Multiple grant numbers should be separated by a comma and space, and where research was funded by more than one agency the different agencies should be separated by a semi-colon, with 'and' before the final funder. Grants held by different authors should be identified as belonging to individual authors by the authors' initials. For example, "This work was supported by the Deutsche Forschungsgemeinschaft (A.B., grant numbers XXXX, YYYY), (C.D., grant number ZZZZ); the Natural Environment Research Council (E.F., grant number FFFF); and the Australian Research Council (A.B., grant number GGGG), (E.F., grant number HHHH)".
}

\backsection[Declaration of interests]{The authors report no conflict of interest.}

% \backsection[Data availability statement]{The data that support the findings of this study are openly available in [repository name] at http://doi.org/[doi], reference number [reference number].}

\backsection[Author ORCID]{
  C.\ J.\ Howland, \url{https://orcid.org/0000-0003-3686-9253};
  C.\ S.\ Ng, \url{https://orcid.org/0000-0002-4643-4192};
  R.\ Verzicco, \url{https://orcid.org/0000-0002-2690-9998};
  D.\ Lohse, \url{https://orcid.org/0000-0003-4138-2255}}

% \backsection[Author contributions]{Authors may include details of the contributions made by each author to the manuscript, for example, ``A.G. and T.F. derived the theory and T.F. and T.D. performed the simulations.  All authors contributed equally to analysing data and reaching conclusions, and in writing the paper.''}

\appendix

\edit{
  \section{Boundary layer transition prediction} \label{app:BL_trans}
  In this appendix, we provide an estimate for the $\Pran$-dependence of the transition to a shear-driven turbulent boundary layer, based on the critical Reynolds number criterion of \citet{landau_fluid_1987}.
  From each simulation, we can calculate a Reynolds number $\Rey_{\delta^\ast}$ based on the displacement thickness $\delta^\ast$ by
  \begin{align}
    \delta^\ast &= \int_0^{x_\mathrm{max}} 1 - \frac{\overline{v}(x)}{V_\mathrm{max}} \,\mathrm{d}x , &
    Re_{\delta^\ast} &= \frac{V_\mathrm{max}\delta^\ast}{\nu} ,
  \end{align}
  where $V_\mathrm{max}$ is the maximum vertical velocity and $x_\mathrm{max}$ is the wall-normal location of this maximum.
  Performing the same linear regression as described in \S\ref{sec:heat_flux} on this data, we obtain the power law relation
  \begin{equation}
    \Rey_{\delta^\ast} = 0.159 \Ray^{0.294} \Pran^{-0.557} .
  \end{equation}
  Assuming (somewhat ambitiously) that this scaling remains valid up to a critical Reynolds number of $\Rey_{\delta^\ast}=\Rey_c = 420$, we infer a $\Pran$-dependent critical Rayleigh number of
  \begin{equation}
    Ra_c = 4.27\times 10^{11} \times \Pran^{1.89} . \label{eq:Ra_crit}
  \end{equation}
  For $\Pran=1$, this gives a value within the transition range of $3.8\times 10^{10}\lesssim Ra_c \lesssim 10^{12}$ predicted by \citet{ng_changes_2017} in figure 10 of that paper.
}

\edit{
  In the context of a melting vertical ice face in the ocean, we can use \eqref{eq:Ra_crit} to estimate the length scales at which a shear-driven boundary layer may be relevant in describing the salt flux towards the ice due to natural convection.
  Although the ice can be considered salt-free, at a water temperature of \SI{2}{\celsius} the interfacial concentration of salinity is approximately \SI{15}{g.kg^{-1}} \citep[see e.g.][]{kerr_dissolution_2015}.
  Combined with an ambient ocean salinity of \SI{35}{g.kg^{-1}}, a haline contraction coefficient of $\beta = \SI{7.86e-4}{(g.kg^{-1})^{-1}}$, a kinematic viscosity of $\nu=\SI{1e-6}{m^2 s^{-1}}$, and a Schmidt number $Sc=\nu/\kappa_S = 2600$, we find
  \begin{equation}
    Ra_c = \frac{g\beta H_c^3 \Delta_S}{\nu\kappa_S} \approx 10^{18} \approx 4\times 10^{14} H_c^3, \hspace{3em} \textrm{implying that} \qquad
     H_c \approx \SI{13.5}{m}.
  \end{equation}
  Note that $H_c$ is the critical \emph{horizontal} length scale.
  In the context of convection at an ice face, where the domain is essentially unbounded in one direction, this is best compared with the local plume width.
  Following \citet{wells_geophysical-scale_2008}, the plume width $H$ can be linearly related to the height $Z$ from the base of the ice by $H\approx 0.1 Z$.
  This relation is based on the constant entrainment rate assumption of classical plume theory as developed by \citet{morton_turbulent_1956}.
  The critical vertical position for a shear-driven boundary layer is then $Z_c=\SI{135}{m}$, associated with a Rayleigh number of $Ra_z=10^{21}$.
  This matches the prediction of \citet{kerr_dissolution_2015} who used GL theory to estimate the transition.
  Over such large vertical distances, other physical phenomena are likely to play an important role in the dynamics, such as ambient stratification \citep{mcconnochie_effect_2016} or the pressure-dependence of the melt condition at the boundary of the ice \citep{hewitt_subglacial_2020}.
  It is therefore unlikely that a shear-driven boundary layer would develop at an ice face \emph{solely} due to natural convection, without some external forcing such as subglacial discharge or a mean horizontal current.
}

\edit{
  \section{Turbulence budgets}
}

\begin{figure}
  \centerline{\includegraphics[width=\linewidth]{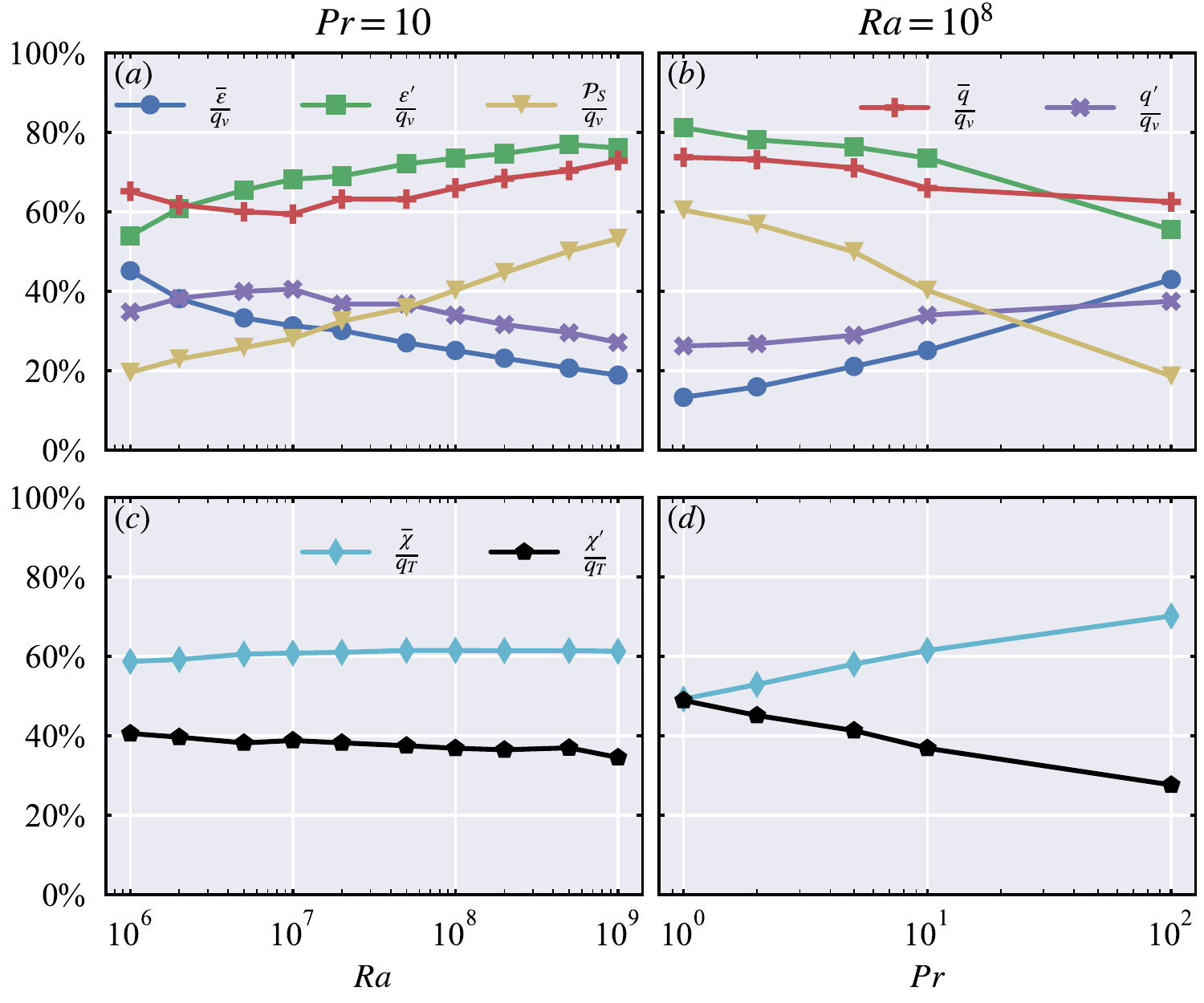}}
  \caption{\edit{
    Relative contributions to the heat flux due to the energy budget terms.
    ($a,b$) Plot of the kinetic energy budget terms as a fraction of the total vertical heat flux; ($c,d$) plot of the thermal dissipation rates as a fraction of the total horizontal heat flux.
    ($a,c$) show variation with Rayleigh number for simulations at fixed $\Pran = 10$, ($b,d$) show variation with Prandtl number for simulations at fixed $\Ray = 10^8$.
  }}
  \label{fig:q_fraction}
\end{figure}

\edit{
  Finally, to gain more insight into the nature of the flow as $\Ray$ and $\Pran$ vary, we present results from the turbulence budgets of our simulations and describe how the turbulent kinetic and thermal dissipation rates are related to the heat flux in the system.
  From the governing equations, we can construct evolution equations for the kinetic energy of the mean flow $\overline{E_K}$, the turbulent kinetic energy $E_K'$, and the equivalent quantities for the temperature field
  \begin{align}
    \overline{E_K} &= \frac{1}{2}\left\langle\left|\overline{\boldsymbol{u}}\right|^2 \right\rangle, &
    {E_K}^\prime &= \frac{1}{2} \left\langle\left|\boldsymbol{u}^\prime\right|^2\right\rangle, &
    \overline{E_T} &= \frac{1}{2}\left\langle |\overline{T}|^2 \right\rangle, &
    {E_T}^\prime &= \frac{1}{2}\left\langle |T^\prime|^2 \right\rangle,
  \end{align}
  where as in the main text an overbar denotes an average over $y$ and $z$, and angle brackets denote a domain average.
  The evolution equations for the kinetic energies read
  \begin{align}
    \frac{d\overline{E_K}}{dt} &= -\mathcal{P}_S - \overline{\varepsilon} + \overline{q}, &
    \frac{d{E_K}^\prime}{dt} &= \mathcal{P}_S - \eps' + q^\prime , \label{eq:KE_evo}
  \end{align}
  where the shear production $\mathcal{P}_S$, which transfers energy between turbulence and the mean flow, is defined as
  \begin{equation}
    \mathcal{P}_S = -\overline{u'\boldsymbol{u}'\cdot \frac{\pd \overline{\bu}}{\pd x}} ,
  \end{equation}
  the dissipation rates of mean KE and TKE are
  \begin{align}
    \overline{\eps} &= \nu \left\langle \left| \frac{\pd \overline{\bu}}{\pd x}\right|^2 \right\rangle , &
    \eps' &= \nu \left\langle \frac{\pd u_i}{\pd x_j} \frac{\pd u_i}{\pd x_j} \right\rangle, 
  \end{align}
  and the vertical heat fluxes due to the mean and turbulent profiles are given by
  \begin{align}
    \overline{q} &= g\alpha \left\langle \overline{v}\overline{T}\right\rangle , &
    q' &= g\alpha \left\langle v'T'\right\rangle .
  \end{align}
  The mean square temperature and the temperature variance evolve according to similar equations, namely
  \begin{align}
    \frac{d\overline{E_T}}{dt} &= -\mathcal{P}_T - \overline{\chi} + q_T, &
    \frac{d{E_T}^\prime}{dt} &= \mathcal{P}_T - \chi' . \label{eq:PE_evo}
  \end{align}
  Here $\mathcal{P}_T$ is an analogous term to the shear production described above, and quantifies the interaction between the mean temperature profile and the turbulent fluctuations:
  \begin{equation}
    \mathcal{P}_T = -\left\langle u'T' \frac{\pd \overline{T}}{\pd x} \right\rangle.
  \end{equation}
  The thermal dissipation rates are given by
  \begin{align}
    \overline\chi &= \kappa \left\langle \left(\frac{\pd \overline{T}}{\pd x} \right)^2 \right\rangle ,&
    \chi' &= \kappa \left\langle \left|\del T\right|^2\right\rangle ,
  \end{align}
  and $q_b$ is the mean horizontal heat flux through the boundaries
  \begin{equation}
    q_T = \frac{\kappa}{2}\left(\left.\frac{\pd \overline{T}}{\pd x}\right|_{x=0} + \left.\frac{\pd \overline{T}}{\pd x}\right|_{x=H}\right) = \frac{\Nuss H}{\kappa \Delta T}. 
  \end{equation}
}

\edit{
  In the statistically steady states reached by our simulations, the energies become constant in time, such that we get the following relations from \eqref{eq:KE_evo} and \eqref{eq:PE_evo}
  \begin{align}
    \overline{q} &= \mathcal{P}_S + \overline{\eps}, &
    \eps' &= \mathcal{P}_S + q', &
    q_T &= \mathcal{P}_T + \overline{\chi}, &
    \mathcal{P}_T &= \chi'.
  \end{align}
  These equations highlight how the total vertical heat flux $q_v = \overline{q} + q^\prime$ can be related to the kinetic energy dissipation rate, and how the horizontal heat flux $q_T$ can be related to the thermal dissipation rate:
  \begin{align}
    q_v &= \overline{q} + q^\prime = \overline{\eps} + \eps^\prime, &
    q_T &= \overline{\chi} + \chi^\prime .
  \end{align}
}

\edit{
  Figure \ref{fig:q_fraction} plots the relative contributions of each of these budget terms to the heat fluxes as a function of $\Ray$ and $\Pran$.
  For the kinetic energy budget terms (shown in panels $a$ and $b$), we observe that the relative contributions of $\eps'$ and $\overline{q}$ increase with $\Ray$ and decrease with $\Pran$.
  This also coincides with an increase in the relative magnitude of the shear production $\mathcal{P}_S$.
  Since $\mathcal{P}_S$ is positive in all our simulations, this means that energy is always (on average) transferred from the mean flow to the turbulent perturbations.
  The trends observed in figure \ref{fig:q_fraction}($a,b$) suggest that the kinetic energy budget terms may be most sensitive to the Reynolds number of the flow.
  By contrast, the relative contributions of the thermal dissipation rates plotted in figure \ref{fig:q_fraction}($c$) show very weak dependence on $\Ray$.
  For $\Pran$ fixed at 10, the dissipation of the mean temperature accounts for 60\% of the horizontal heat flux, and this fraction changes by less than 3\% over three decades of $\Ray$.
  As $\Pran$ increases the relative contribution of $\overline{\chi}$ becomes greater.
  This highlights once again the key role that the thin, conductive boundary layers, whose strong gradients contribute to $\overline{\chi}$, have on the heat flux in vertical convection at high $\Pran$.
}

\bibliographystyle{jfm}
\bibliography{VC}

\end{document}